\documentclass[twocolumn, times, tighten,twocolappendix, a4paper]{aastex701}
\usepackage{amsmath,amssymb}
\usepackage{txfonts}
\usepackage[T1]{fontenc}
\usepackage{graphicx,multirow,hyperref,url,color,xspace}

\newcommand{\g}{$\gamma$}
\newcommand{\msun}{{M}_{\sun}}

\newcommand{\nustar}{{NuSTAR}\xspace}
\newcommand{\nicer}{NICER\xspace}

\newcommand{\source}{{MAXI J1631--479}\xspace}

\newbox\grsign \setbox\grsign=\hbox{$>$} \newdimen\grdimen \grdimen=\ht\grsign
\newbox\simpropbox
\setbox\simpropbox=\hbox{\raise.5ex\hbox{$\propto$}\llap
 {\lower.5ex\hbox{$\sim$}}}\ht2=\grdimen\dp2=0pt

\begin{document}

\title{The strong Fe K line and spin of the black-hole X-ray binary MAXI J1631--479}

\author[0000-0002-0333-2452]{Andrzej A. Zdziarski}
\affiliation{Nicolaus Copernicus Astronomical Center, Polish Academy of Sciences, Bartycka 18, PL-00-716 Warszawa, Poland} 
\email[show]{aaz@camk.edu.pl}

\author[0000-0003-3499-9273]{Swadesh Chand}
\affiliation{Institute of Astronomy, National Tsing Hua University, Hsinchu 300044, Taiwan}
\affiliation{Inter-University Centre for Astronomy and Astrophysics, Pune 411007, India}
\email{swadesh.chand@iucaa.in}

\author[0000-0003-1589-2075]{Gulab Dewangan}
\affiliation{Inter-University Centre for Astronomy and Astrophysics, Pune 411007, India}
\email{gulabd@iucaa.in}

\author[0000-0002-7609-2779]{Ranjeev Misra}
\affiliation{Inter-University Centre for Astronomy and Astrophysics, Pune 411007, India}
\email{rmisra@iucaa.in}

\author[0000-0001-7606-5925]{Micha{\l} Szanecki}
\affiliation{Faculty of Physics and Applied Informatics, {\L}{\'o}d{\'z} University, Pomorska 149/153, PL-90-236 {\L}{\'o}d{\'z}, Poland}
\email{mtszanecki@gmail.com}

\author[0000-0002-8231-063X]{Bei You}
\affiliation{School of Physics and Technology, Wuhan University, Wuhan 430072, China}
\email{youbei@whu.edu.cn}

\author[0009-0003-8610-853X]{Maxime Parra}
\affiliation{Department of Physics, Ehime University, 2-5, Bunkyocho, Matsuyama, Ehime 790-8577, Japan}
\email{maximeparrastro@gmail.com}

\author[0000-0003-1780-5641]{Gr{\'e}goire Marcel}
\affiliation{Department of Physics and Astronomy, FI-20014 University of Turku, Finland}
\email{gregoiremarcel26@gmail.com}

\begin{abstract}
We study the transient black hole binary MAXI J1631--479 observed simultaneously by NICER and NuSTAR in its soft spectral state. Its puzzling feature is the presence of a strong and broad Fe K line, while the continuum includes a prominent disk blackbody and a very weak power-law tail. The irradiation of the disk by a power-law spectrum fitting the tail is far too weak to explain the strong line. Previous proposals included the idea that the Fe K emission is intrinsic to the disk. Here, we propose that the strong line can be explained by the irradiation of the disk by photons from Comptonization of the disk blackbody by coronal electrons. One crucial effect is that the shape of the irradiating spectrum at $\lesssim$10 keV reflects that of the disk blackbody; it is strongly curved and has a higher flux than what would be produced by a fit with a power-law irradiation. The other effect is a relativistic enhancement of the backscattered coronal flux incident on the disk. Both effects together can account for the line, although the latter is modeled only quantitatively. While this result is independent of the physical model used for disk emission, the fitted spin depends heavily on that model. When employing a Kerr disk model for a thin disk with color correction, the fitted spin appears retrograde, rare for a Roche-lobe overflow binary. A model that accounts for both the finite thickness of the disk and radiative transfer yields a spin of $a_*\approx0.8$--0.9. 
\end{abstract}

\section{Introduction} \label{intro}

\source is a transient X-ray binary (XRB), which went into outburst in late 2018 and was detected by MAXI \citep{Kobayashi18} and NuSTAR \citep{Miyasaka18}. The donor has not yet been detected, indicating that the system is a low-mass X-ray binary (LMXB). There is no Gaia counterpart, and, given the strong extinction toward the source, $E(B$--$V)\approx 4.7$ in the BlackCAT catalog\footnote{\url{https://www.astro.puc.cl/BlackCAT/}; \citet{Corral-Santana16}.}, a detection of the donor may be impossible. Its nature as a transient black hole (BH) LMXB is supported by its timing and spectral properties \citep{Xu20_J1631, Rout21, Bu21, Fu22}, which are typical of those sources.

Its BH mass, $M$, and the distance, $D$, remain uncertain. \citet{Rout23} claimed a very high BH mass, $>26\msun$. This was implied by the disk blackbody model seen at high inclination, $i\approx 71\pm 1\degr$, found by them. On the other hand, low inclination values were found in the spectral studies by \citet{Xu20_J1631}, $i=29\pm 3\degr$, and by \citet{Draghis24}, $i=22^{+10}_{-12}\degr$. Its dimensionless spin parameter was estimated to be very high, $a_*>0.94$ \citep{Xu20_J1631}, $a_*>0.996$ \citep{Rout23}, and $a_*=0.95^{+0.04}_{-0.08}$ \citep{Draghis24}. Those studies used the X-ray reflection method \citep{Bambi21} in the soft spectral states of \source. However, the results of this method crucially depend on the shape of the irradiating continuum, which is likely to be different from the usually assumed power law shape \citep{Zdziarski25a}, especially in the soft state, when it is apparently formed by Compton upscattering of the disk emission by energetic electrons. 

Here, we measure the spin using the continuum method \citep{McClintock14} jointly with the reflection method, which is applicable in the soft spectral state \citep{Parker16}. The continuum method relies on the dependence of the innermost stable circular orbit (ISCO) radius on spin \citep{Bardeen72} and uses relativistic accretion disk models to fit observed X-ray spectra. We study a set of simultaneous NICER and NuSTAR observations of this source in the very soft state, i.e., ones with a strong disk blackbody and a very weak high-energy tail. We use two models of relativistic disks, {\tt kerrbb} \citep{Li05} and {\tt slimbh} \citep{Straub11}. 

\begin{table*}[t!]
\centering
\caption{The log of the observations \label{log}}
\renewcommand{\arraystretch}{0.95}
  \begin{tabular}{cccccccc}
    \hline
NICER           & Start Time &  Exposure (s) & NuSTAR & Start Time  & Exposure (s) & Exposure (s)\\
Obs. ID         &  End Time  &  & Obs. ID & End Time & FPM A & FPM B\\
 \hline

 1200500103 & 2019-01-17 04:37:08 & 3016 & 90501301001 & 2019-01-17 02:38:58 & 6670 & 7018 \\
Part I           & 2019-01-17 12:28:51 &      &    Part I         & 2019-01-17 13:17:04 &       &      \\
\hline
    1200500103        & 2019-01-17 16:58:30 & 4870 &    90501301001   & 2019-01-17 15:55:37 & 7471  & 7949 \\
 Part II           & 2019-01-17 23:24:54 &      &  Part II           & 2019-01-18 05:23:26 &       &      \\
\hline
 \end{tabular}
 \end{table*}

A puzzling feature of this source is the presence of a strong and broad Fe K emission line feature, with EW $\approx 180$--210 eV \citep{Xu20_J1631}, in the presence of a weak high-energy tail. When the tail is fitted by a power law plus reflection, the latter is way too weak to account for that feature, especially in the initial part of the studied observation \citep{Xu20_J1631}. The initial solution proposed by \citet{Xu20_J1631} was that the broad line is intrinsically emitted by the disk, being formed due to the transfer of the blackbody emission from deep in the disk to its surface. However, this contradicts standard disk spectra that account for vertical radiative transfer (\citealt{Davis05, Davis06}; S. Davis, private communication), which do not exhibit strong Fe K features. On the other hand, there is an ongoing debate on the applicability of the standard disk model to the soft state of BH XRBs (which we discuss in Section \ref{discussion}), with the modified models still lacking detailed spectral predictions. Then, the Fe K line might still be intrinsic to the accretion disk in this source, as proposed by \citet{Xu20_J1631}.

Later, \citet{Rout23} proposed that the Fe K feature is due to the reflection of blackbody radiation emitted by the disk and returning to it in strong gravity \citep{Cunningham75,  Schnittman09}. However, the model of \citet{Rout23} included the disk blackbody, the irradiating blackbody (dominant), and its reflection in their fits. However, the irradiating blackbody should not be included at all as a separate component since it is already accounted for by the disk blackbody. 

In the present work, we propose a new solution, in which the Fe K feature is still due to the reflection of the coronal radiation due to Compton scattering of the disk blackbody by relativistic electrons. That reflection can be significantly stronger than that from a power-law spectrum fitted to the high-energy tail due to both the strong concave curvature of the actual spectrum from Comptonization, and the relativistic enhancement of the emission toward the disk with respect to that outside \citep{Ghisellini91}. 

\section{The data and variability}
\label{data}

\begin{figure}[t!]
\centerline{\includegraphics[width=0.9\columnwidth]{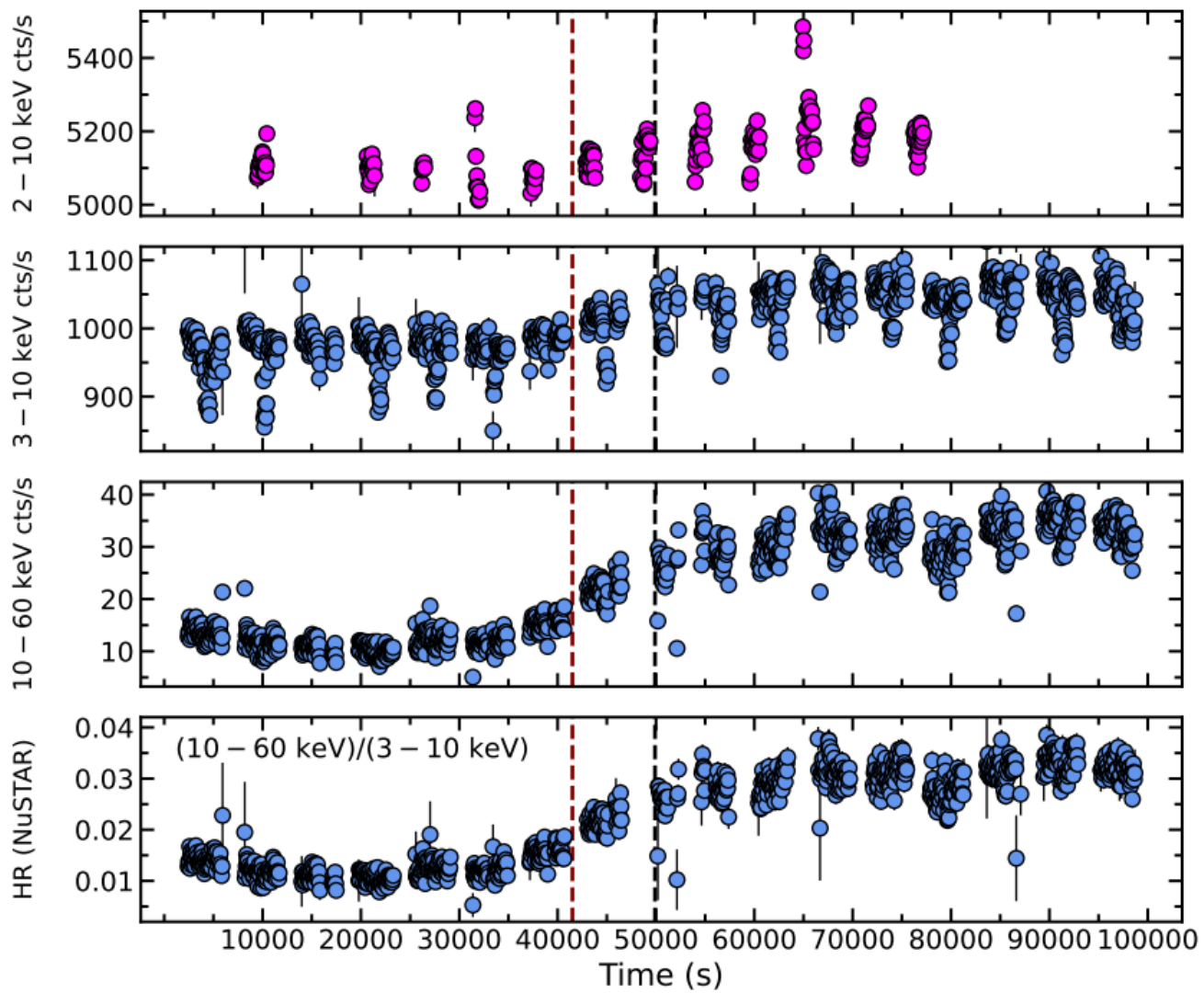}}
\caption{The NICER 2--10 keV and NuSTAR 3--10 and 10--60 keV light curves, and the NuSTAR hardness ratio during the studied observations. The zero time corresponds to the start of the NuSTAR observation, and the vertical dashed lines show the end of Part I (red) and the beginning of Part II (black; see Table \ref{log}). 
}\label{lc}
\end{figure}

The MAXI \citep{Matsuoka09} and Swift/BAT \citep{BAT} light curves of the outburst are shown in Figure 1 of \citet{Xu20_J1631}. We study here the \nustar observation denoted as OBS1 in that paper (ObsID 90501301001), which was performed on MJD 58500, after the peak of the MAXI light curve and before the peak of the BAT light curve. Its spectra consist of a strong disk blackbody component and a weak high-energy tail. We define here Part I of the OBS1 as in Figure 1 and Table 1 of \citet{Xu20_J1631}. However, we define here Part II excluding a short transitional period after Part I (between the vertical dashed curves in Figure \ref{lc}). In addition, we consider simultaneous NICER observations within the same epochs, as presented in \citet{Rout21}.  The log of the observations studied here is given in Table \ref{log}, and the light curves of the detectors and the NuSTAR hardness ratio are shown in Figure \ref{lc}.  

The \nustar data are used in the 3--79 keV range. We processed them using the standard task \texttt{nupipeline} in the NuSTARDAS package, using the option for bright sources\footnote{{\tt statusexpr=\\"(STATUS==b0000xxx00xxxx000)\&\&(SHIELD==0)"}.}. Source and background spectra were extracted using the task \texttt{nuproducts} from circular regions of $100\arcsec$ radius, centered on the source position and from a source-free region, respectively. The corresponding response matrix files and auxiliary response files were generated simultaneously. Despite the \nustar and NICER exposures being nearly simultaneous, we find the average spectrum from NICER shows some disagreement with the NuSTAR average spectrum in the overlapping energy range (see Section \ref{spectral}). During the reduction pipeline of NICER, the script {\tt nicerl3-spect} added 1.5\% systematic error\footnote{\url{https://heasarc.gsfc.nasa.gov/lheasoft/ftools/headas/nicerl3-spect.html}} to the detector channels in the 0.5--9.0\,keV range, and more (up to 2.5\%) above 9 keV. In addition, the NICER spectrum shows a strong positive feature between 1 and 2 keV, which appears to be due to a calibration issue (see, e.g., \citealt{Hall25}). Thus, we use the NICER spectrum in the 2--10 keV energy range. The spectral data have been optimally binned \citep{Kaastra16}, with the additional requirement that each bin contain at least 25 counts. 

We have also calculated the fractional variability of the light curves, using both the NICER and NuSTAR data. The results for the frequency range of 0.01--1 Hz are shown in Table \ref{rms}. We see that the data up to 15 keV show very weak variability. The accretion disk spectrum dominates this range. On the other hand, the high-energy tail exhibits strong variability, particularly in Part I. 

\section{Spectral fits}
\label{spectral}

\subsection{The method}
\label{method}

We follow the self-consistent soft-state spectral modeling developed by \citet{Zdziarski25a}. The main new feature of this method is the use of convolution for both calculating the spectrum of Comptonized disk emission and the reflection spectrum from the former spectrum irradiating the disk. This avoids the assumption that the incident spectra are power laws with either exponential or thermal Compton cutoffs, inherent in either {\tt relxill} or {\tt relxillCp} models \citep{GK10, Garcia18, Dauser16}. 

\begin{table}
\centering
\caption{The fractional rms variablity in the 0.01--1 Hz frequency range \label{rms}}
\vskip -0.4cm
  \begin{tabular}{cccccccc}
    \hline
Detector & Energy range & rms Part I & rms Part II\\  
\hline
NICER & 2--10 keV & $0.65\pm 0.08\%$ & $0.62\pm 0.06\%$\\
NuSTAR & 3-15 keV & $0.64\pm 0.05\%$ & $0.60\pm 0.06\%$\\
NuSTAR & 15-25 keV & $16.4\pm 3.3\%$ & $4.4\pm 3.4\%$\\
\hline
 \end{tabular}
 \end{table}

For Comptonization, we use a convolution version of the {\tt compps} model of \citet{PS96}, {\tt comppsc}\footnote{\url{https://github.com/mitsza/compps_conv}}. The reflection spectrum of the emission incident on the disk is modeled by the convolution model {\tt xilconv} \citep{Kolehmainen11, Garcia13, MZ95}, in which the reflection strength, ${\cal R}$, is defined as a fraction with respect to the case with an isotropic source above a slab (${\cal R}=1$). The reflected spectrum is then relativistically broadened using {\tt relconv} \citep{Dauser10}. The values of the inclination and the spin are linked between the disk and reflection models. Comptonization is characterized by the electron temperature in energy units, $kT_{\rm e}$, and the Thomson optical depth, $\tau_{\rm T}$. As found by \citet{Zdziarski25a}, the differences in the main fitted parameters between the slab and spherical geometries of {\tt comppsc} are small, and we thus use the latter using the (fast) option {\tt geom=0}. 

High-energy tails in the soft state, when modeled by Comptonization, require the presence of nonthermal tails beyond the Maxwellian electron distribution, in particular in Cyg X-1 \citep{Zdziarski24b} and GX 339--4 \citep{Zdziarski25a}. The tail used in {\tt comppsc} is a power law in the electron momentum, i.e., $(\gamma\beta)^{-p}$ (where $\beta$ is the dimensionless electron velocity) from $\gamma_{\rm min}$, at which the normalizations of the Maxwellian and of the power law are equal, up to $\gamma_{\rm max}$. In our fits, we obtain relatively large values of $p$, for which the fits are independent of $\gamma_{\rm max}$, which we thus set equal to 100.

We use two models of relativistic disks. One is {\tt kerrbb} \citep{Li05}, which models geometrically thin disks using the formalism of \citet{NT73} and assumes a local spectrum to be a blackbody with a color correction, $f_{\rm col}$, which we assume to equal 1.7 \citep{Davis19}. The other is {\tt slimbh} (described in \citealt{Straub11}), based on the calculations of \citet{Sadowski09, Sadowski11a} and \citet{Sadowski11b}. It accounts for the finite disk thickness, which is important for $L\gtrsim 0.1 L_{\rm E}$ \citep{Abramowicz88}. Furthermore, it treats the local spectra using the radiative-transfer calculations of \citet{Davis05} and \citet{Davis06} when the option $f_{\rm col}=-1$ is set. These spectra depend on the viscosity parameter, which we set as $\alpha=0.1$.

Our model for spectral fitting in {\sc xspec} \citep{Arnaud96} is
\begin{align}
&{\tt plabs*tbfeo*comppsc}\{{\tt disk}+\nonumber\\
&{\tt relconv[xilconv(comppsc}({\tt disk}))]\},
\label{xspec_model}
\end{align}
where {\tt disk} = {\tt kerrbb} or {\tt slimbh}. The first term in the curly brackets gives the disk emission alone, and the second term gives the relativistically broadened reflection. Both are subject to Comptonization, given by the {\tt comppsc} term outside the brackets. 

Then, {\tt plabs} accounts for small differences in the spectral slope between the NuSTAR A and B units and the NICER, which we found clearly present in the data. We note that the standard {\sc xspec} model {\tt plabs} is equivalent to the external model {\tt CRABCORR}, defined by \citet{Steiner10} for this purpose and widely used. Both multiply the model spectra by $K E^{-\Delta\Gamma}$, where $K=1$ and $\Delta\Gamma=0$ for the \nustar A unit, while they are free for NuSTAR B and NICER. We found that requiring $\Delta\Gamma=0$ for NuSTAR B increases the $\chi^2$ of the fits by $\sim$120. The interstellar absorption is modeled by {\tt tbfeo} \citep{Wilms00}, where we assume the abundances of \citet{Wilms00}. We have found that while the O relative abundance is compatible with unity (which we hereafter set), allowing the Fe abundance to be free reduces $\chi^2$ by $\sim$10. Convolution models require the energy grid to be substantially larger (on both sides) than the detector's grid, and we thus set it using the command {\tt energies 0.01 1000 2000 log}. We use 90\% confidence uncertainties ($\Delta\chi^2 \approx 2.71$; \citealt{Lampton76}). 

\subsection{Individual fits}
\label{ind}

\begin{figure*}[t!]
\centerline{\includegraphics[width=6.5cm]{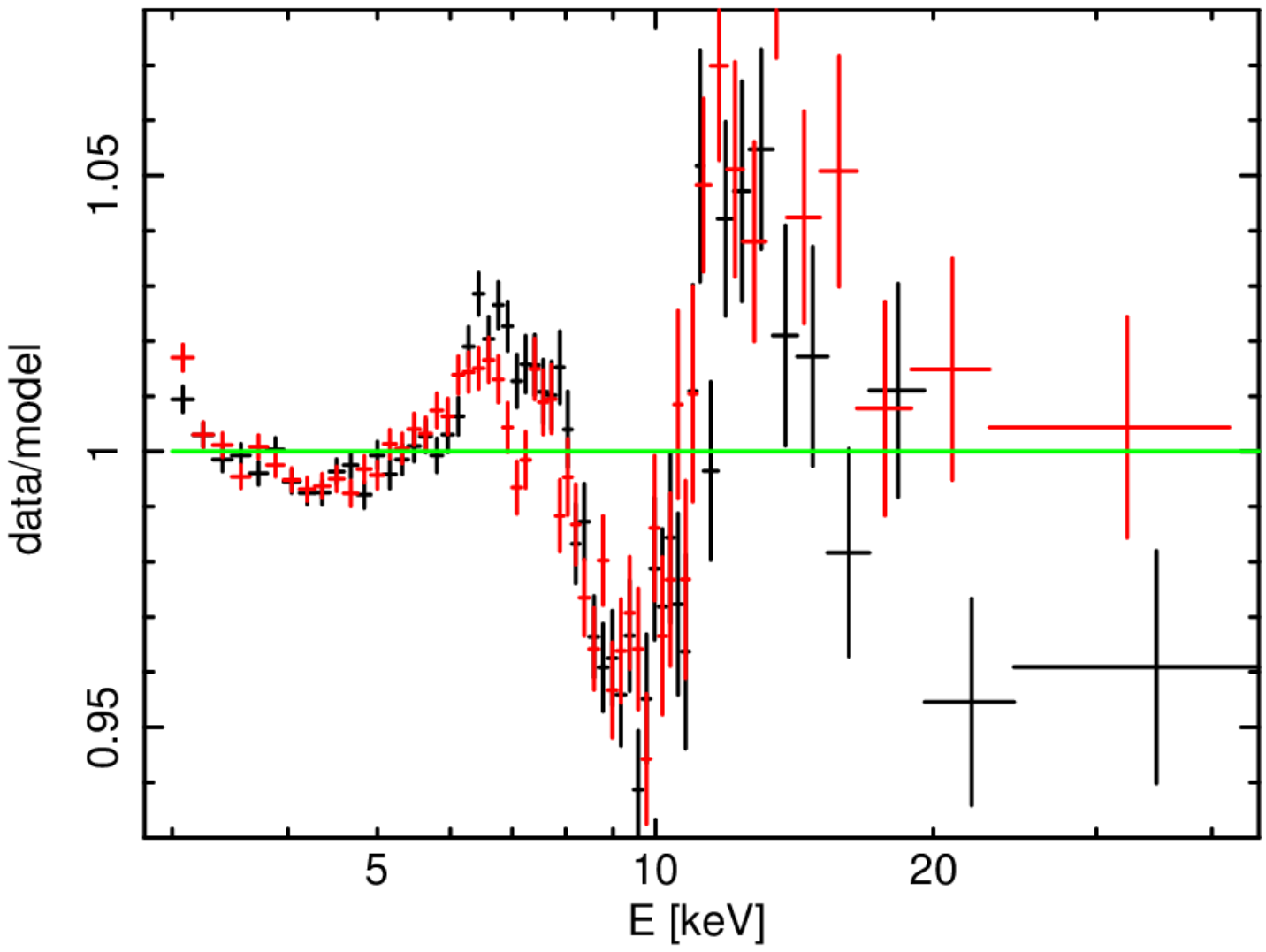}\includegraphics[width=6.5cm]{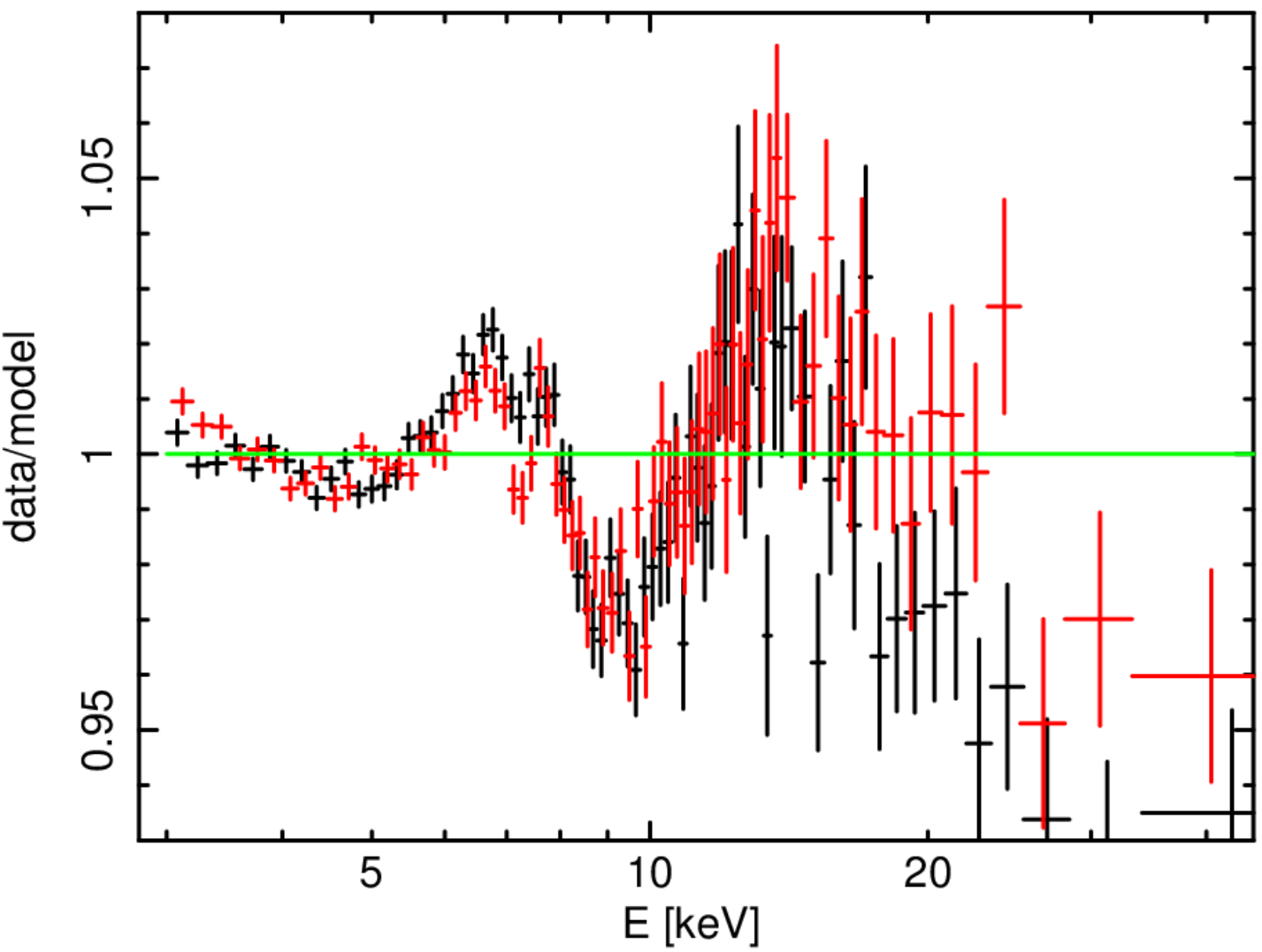}}
\caption{The data to model ratios for the model fitted with {\tt slimbh} to the NuSTAR data alone and without the reflection component for Part I (left) and Part II (right). We observe patterns characteristic of Compton reflection, as well as a marked difference in the spectra from the two units around 7 keV.
}\label{ratio}
\end{figure*}

We initially fitted the spectra of Parts I and II separately. We first demonstrate the presence of Fe K residuals and reflection using individual data. We fit the NuSTAR data alone without reflection with {\tt slimbh}, i.e., using Equation \ref{xspec_model} without the second term in the curly brackets. We obtain poor fits with $\chi^2_\nu=1120/298$ and 1087/391, respectively, with strong residuals indicating significant reflection. The data-to-model ratios are shown in Figure \ref{ratio}. We see distinct Fe K features, edges at energies above them, and reflection humps at $E\gtrsim 10$ keV, confirming the result of \citet{Xu20_J1631}. However, we observe a clear discrepancy in both Parts I and II of the profiles in the 7.00--7.37 keV range between the A and B units, with the latter showing clear dips. These dips were attributed to the presence of an absorption line by \citet{Rout23}. On the other hand, this issue was discussed in detail in \citet{Draghis24}, who attributed it to the Sun's proximity, which destabilizes source tracking, and to NUSTAR B being partly in the detector gap. They attempted to improve data extraction by implementing a custom procedure. However, the marked difference between the A and B profiles around 7 keV remained virtually unchanged, as shown in their Figure 19. Therefore, we kept using the standard data extraction, but we excluded the 7.00--7.37 keV from the NuSTAR B. In that case, we find full agreement between the two units (after applying the small-slope correction; see Table \ref{fits}).

We fit the NICER+NuSTAR data with {\tt slimbh}, including the reflection. We obtain $\chi_\nu^2$ values of 493/419 and 602/503, respectively. We find well-constrained and similar values of the spin parameter, $a_*=0.88_{-0.03}^{+0.06}$ and $0.83_{-0.06}^{+0.04}$ for Part I and II, respectively. The fitted spins, masses, inclinations, and distances are also relatively similar. 

We then use {\tt kerrbb} with reflection. We obtain $\chi_\nu^2$ values of 511/419 (significantly higher than that for the model with {\tt slimbh}) and 604/503 for Parts I and II, respectively. The spin is only weakly constrained, $a_*=-0.60_{-0.24}^{+0.33}$ and $-0.26_{-0.74}^{+1.02}$ for Parts I and II, respectively. 

\subsection{Joint fitting}
\label{joint}

\begin{figure*}[t!]
\centerline{\includegraphics[width=0.9\columnwidth]{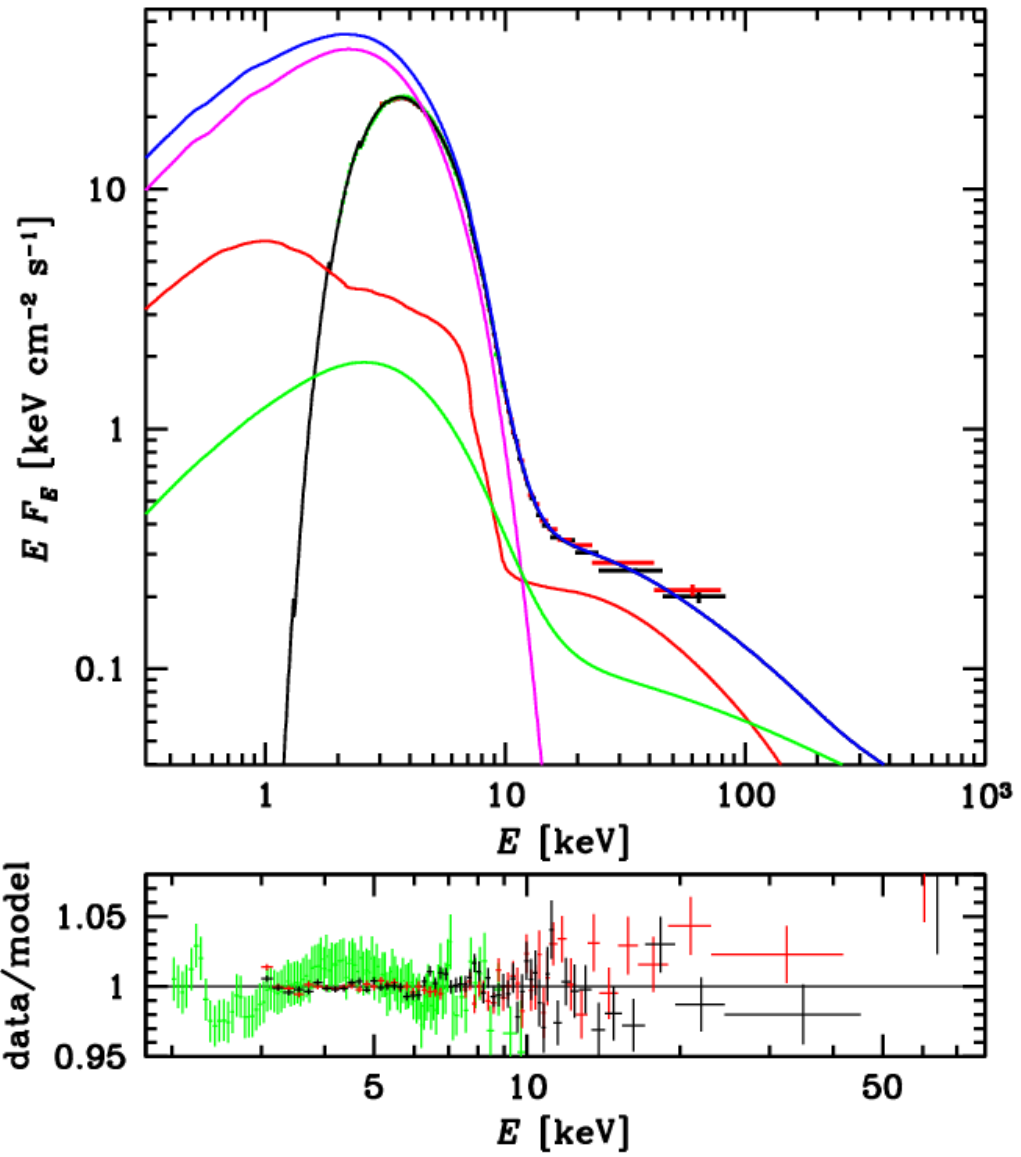}\includegraphics[width=0.9\columnwidth]{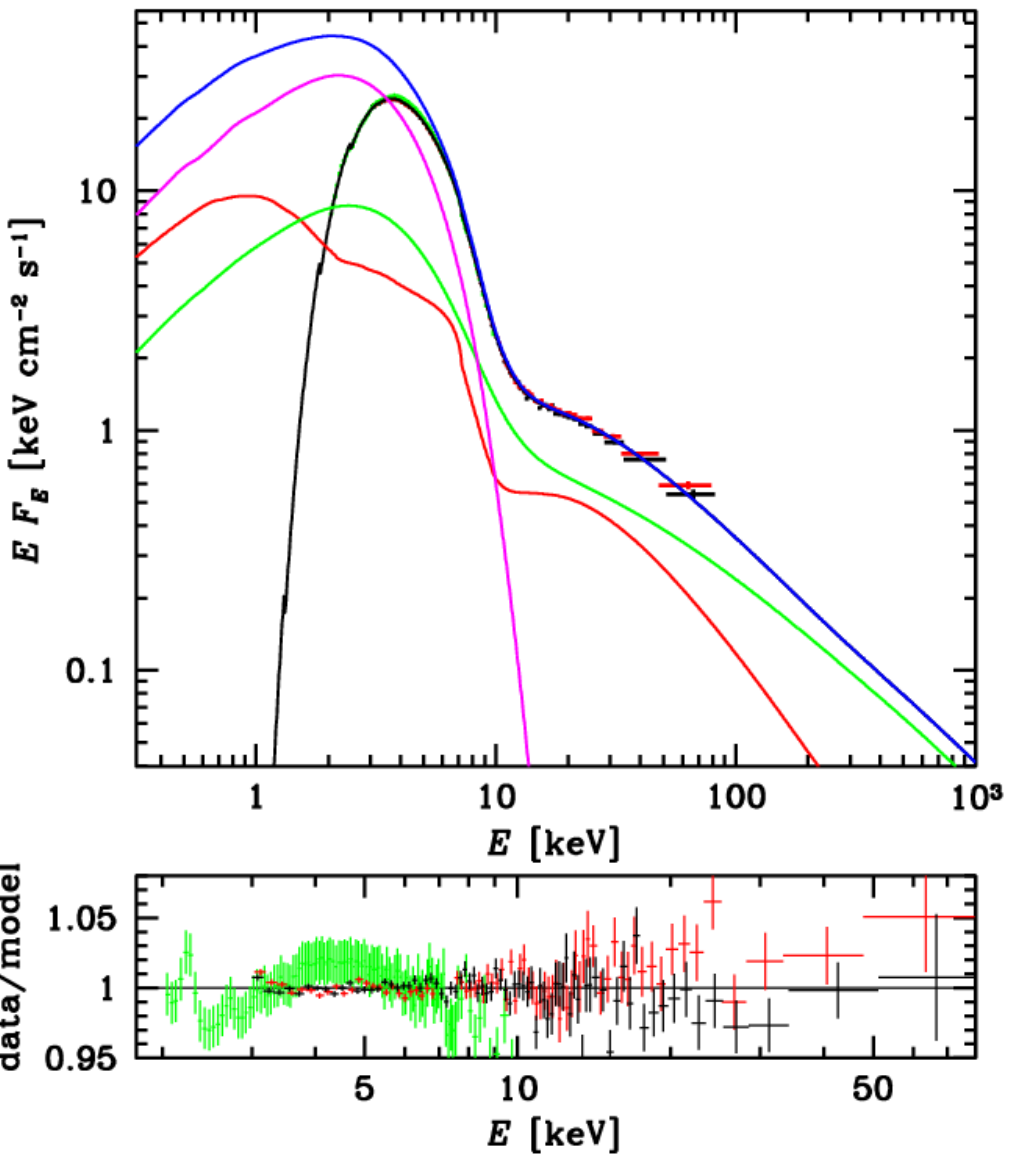}}
\caption{The \nicer (green) and \nustar A and B (black and red, respectively) unfolded spectra (top panels) and data-to-model ratios (bottom panels) for the joint fit using {\tt slimbh}. The Part I and II results are shown in the left and right panels, respectively. The total model spectra and the unabsorbed ones are shown by the solid black and blue curves, respectively. The unscattered disk emission, scattering alone, and reflection are shown by the magenta, green, and red curves, respectively. Their sums equal the total unabsorbed spectra (in blue).
}\label{spectra}
\end{figure*}

\begin{table}
\caption{The results of the joint spectral fitting with {\tt slimbh}}
\vskip -0.4cm
   \centering\begin{tabular}{lccc}
\hline
Component & Parameter & Part I & Part II \\
\hline
Disk & $L_{\rm disk}/L_{\rm E}$ & $0.26^{+0.01}_{-0.01}$ & $0.25^{+0.03}_{-0.03}$ \\
\hline 
Compton  & $kT_{\rm e}$\,[keV]   & $18^{+4}_{-4}$ & $8.2^{+1.9}_{-1.0}$  \\
  & $\tau$    & $0.25^{+0.30}_{-0.10}$ & $0.68^{+0.29}_{-0.18}$  \\
  & $p$    & $3.7_{-0.1}^{+0.2}$    & $4.4_{-0.1}^{+0.2}$\\
  & $\gamma_{\rm min}$&  $1.29^{+0.08}_{-0.02}$   & $1.12^{+0.02}_{-0.03}$  \\
  &$f_{\rm cov}$ & $0.28^{+0.23}_{-0.10}$ & $0.61^{+0.12}_{-0.12}$    \\
\hline
Reflection & ${{\cal R}}$ & $2.8^{+1.6}_{-1.0}$  & $1.0_{-0.3}^{+0.3}$  \\
  &$\log_{10}\xi$    & $4.3_{-0.1}^{+0.2}$ & $4.3_{-0.2}^{+0.1}$  \\
& $q$    & $4.0^{+0.2}_{-0.2}$     & $3.7^{+0.1}_{-0.2}$  \\
\hline
Joint parameters & $N_{\rm H}$ $[10^{22}$\,cm$^{-2}]$ & \multicolumn{2}{c}{$6.8^{+0.7}_{-0.9}$} \\
  & $Z_{\rm O,ISM}$ & \multicolumn{2}{c}{1f}  \\
  & $Z_{\rm Fe,ISM}$&  \multicolumn{2}{c}{$0.50^{+0.17}$}  \\    
  & $M\,[\msun]$& \multicolumn{2}{c}{$10.7_{-0.9}^{+1.5}$}\\
  & $D\,$[kpc] &  \multicolumn{2}{c}{$5.1^{+1.1}_{-0.4}$}\\ 
  & $a_*$& \multicolumn{2}{c}{$0.86^{+0.03}_{-0.03}$}    \\
  & $i\,[\degr]$ &  \multicolumn{2}{c}{$36^{+2}_{-2}$} \\ 
  & $Z_{\rm Fe,disk}$  & \multicolumn{2}{c}{$4.6_{-0.6}^{+1.4}$}\\  
  & $\gamma_{\rm max}$&  \multicolumn{2}{c}{$100$f}  \\
\hline 
Cross-calibration &$\Delta\Gamma_{\rm NICER}$&  \multicolumn{2}{c}{$-0.11_{-0.01}^{+0.01}$}\\
& $K_{\rm NICER}$ &$0.84_{-0.01}^{+0.01}$ &$0.85_{-0.01}^{+0.01}$\\
&$\Delta\Gamma_{\rm NuSTAR\,B}$& \multicolumn{2}{c}{$0.017_{-0.002}^{+0.003}$}\\
&$K_{\rm NuSTAR\,B}$ & $1.02_{-0.01}^{+0.01}$ & $1.01_{-0.01}^{+0.01}$\\
\hline
  & $\chi_\nu^2$  &  \multicolumn{2}{c}{1108/931} \\
\hline
\end{tabular}
\tablecomments{$L_{\rm disk}$ is the unabsorbed disk luminosity; $L_{\rm E}$ is the Eddington luminosity; $\xi\equiv 4{\pi}F_{\rm{irradiating}}/n$ is the ionization parameter; $n$ is the electron density; $q$ is the power law index of the radial dependence of the emissivity used in {\tt relconv}. $Z_{\rm O,ISM}\geq 0.5$ was assumed. }
\label{fits}
\end{table}

Here, we jointly fit the two data sets. We assume common values of the interstellar absorption parameters, mass, distance, inclination, spin, and the Fe abundance of the reflecting medium. We first use {\tt kerrbb}. We obtain $\chi_\nu^2= 1129/931$ giving a negative spin of $a_*= -0.40^{+0.15}_{-0.23}$. We have searched the parameter space for a solution with $a_*>0$, but none has been found. 

Then, we use {\tt slimbh}, which yields a better fit with $\chi_\nu^2= 1108/930$. It gives a well-constrained, positive, and relatively high spin of $a_*= 0.86^{+0.03}_{-0.03}$. It is also a more physically motivated model, incorporating the dependence on the Eddington ratio through both the disk scale height and the radiative transfer effects \citep{Straub11}. We then selected {\tt slimbh} as our final model. See Table \ref{fits} for all the parameters. The spectra and residuals from the joint fit are shown separately for Parts I and II in Figure \ref{spectra}. We have also tested the effect of not including the correction for the spectral slope between the NuSTAR A and B units. When setting $\Delta \Gamma_{\rm NuSTAR\,B}=0$, we obtain a significantly worse fit, $\Delta \chi^2=+119$, but the spin remains the same with a slightly lower uncertainty, $a_*= 0.86^{+0.03}_{-0.02}$.

The corona covers roughly half of the disk. The obtained mass is within the range found for BHs in LMXBs \citep{Ozel10}. We note that {\tt slimbh} does not allow for $a_*<0$. However, the $\chi^2$ dependence on the decreasing $a_*$ is approximately monotonically increasing, and $\Delta\chi^2=+45$ at $a_*=0$. Thus, the presence of a local minimum at $a_*<0$ appears unlikely. An essential feature of the spectra shown in Figure \ref{spectra} is that the scattered disk emission (green), which is both emitted outside and impinges on the disk, has a shape very different from a power law. We observe no significant residuals in the \nustar data, but the \nicer spectrum shows wavy residuals, indicative of instrumental issues. The apparent line above 2 keV is due to the gold edge of the NICER detector. Since its contribution to $\chi^2$ is small, we opted not to model it. 

Finally, we consider a model with a warm, optically thick corona on top of the accretion disk, as previously applied to Cyg X-1, LMC X-1, M33 X-7, and GX 339--4 \citep{Belczynski24, Zdziarski24a, Zdziarski24b, Zdziarski25a}. In those cases, the presence of such coronae allowed spin values close to null, and significantly improved the fit for Cyg X-1. Such coronae were modeled there by optically thick thermal Comptonization using the code {\tt thcomp} \citep{Z20_thcomp}. The differences of this model with respect to that of Equation \ref{xspec_model} is the addition of the {\tt thcomp} factor in front of the {\tt disk} factor and that a free color correction in {\tt slimbh} is used instead of the model including radiative transfer (see a discussion in \citealt{Zdziarski25a}). However, we have not identified a model with a warm corona that yields a better fit than our main model. The best fit has $a_*=0.00^{+0.08}$ and $\chi_\nu^2= 1120/926$. Thus, we conclude that while such a component may still be present in \source, there is no statistical evidence for it.

\section{Discussion}
\label{discussion}

\begin{figure}[t!]
\centerline{\includegraphics[width=\columnwidth]{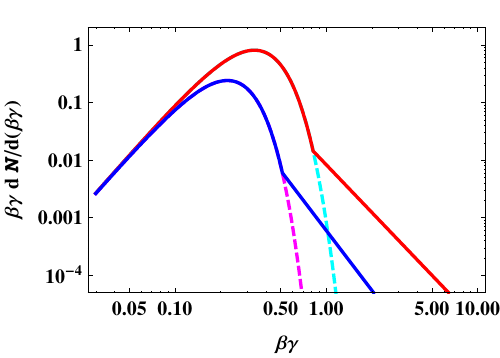}}
\caption{The hybrid electron distributions corresponding to the joint best fit using {\tt comppsc} and {\tt slimbh}, see Figure \ref{spectra} and Table \ref{fits}. The distributions for Part I and II are shown by the red and blue solid curves, respectively. The dashed cyan and magenta curves show the corresponding Maxwellian spectra. The normalization is arbitrary.
}\label{hybrid}
\end{figure}

The main goal of this study is explain the strength of the Fe K line, which was found before to be stronger than that expected from a power-law emission of the corona irradiating the disk. In our approach, we accounted for the incident spectra differing from a power law, assuming that the irradiating spectrum results from Comptonization of the disk emission. The Comptonization of disk blackbody photons (as well as reflection) is modeled self-consistently using convolution models. The scattered spectrum is found to be strongly curved, with the shape far from a power law, see the green curve in Figure \ref{spectra}. To quantify this difference, we have fitted a model with {\tt slimbh+relxill} to the data. {\tt relxill} calculates relativistic reflection assuming the incident spectrum is an e-folded power law (setting the e-folding energy to 1000 keV). The fluxes of this spectrum at the Fe K edge at 7 keV were found in our model to be a factor of $\approx$2 times higher than the flux of the power-law fitted with reflection to the high-energy tail. Thus, the irradiating flux is enhanced by that factor, which helps explain the relatively strong Fe K line observed in the spectrum. This also demonstrates that spectral fits assuming the incident spectrum is a power law, such as {\tt relxill} or {\tt reflionx}, would yield incorrect results for soft-state sources with high-energy tails. Instead, we advocate the use of convolution models.

The fitted reflection fractions are $2.8^{+1.6}_{-1.0}$ and $1.0_{-0.3}^{+0.3}$ for Parts I and II, respectively. This difference is shown in Figure \ref{spectra}, where the reflection spectrum (red) is compared with the Comptonized irradiating spectrum (green). The relatively high values for Part I can then be explained by the effect of anisotropy of a slab corona containing relativistic electrons \citep{Ghisellini91}. The seed blackbody photons from the disk are emitted upward and are scattered by the electrons in the corona. The highest probability of collision occurs when electrons move downward. When the electrons are relativistic, they scatter preferentially forward, thus in the direction of the disk. This leads to an enhancement of the reflection when measured relative to the radiation emitted toward the observer. The enhancement can be large, even by a factor of several \citep{Ghisellini91}. The higher value for Part I is likely due to its coronal electrons being more relativistic than those of Part II. This is illustrated in Figure \ref{hybrid}, which shows the fitted hybrid (Maxwellian followed by a power law) distributions. The hybrid electrons in our calculations have distinct high-energy tails, and a part of them is relativistic, especially for Part I with a large fraction of electrons with $\beta\gamma\gtrsim 1$, as shown in Figure \ref{hybrid}. Thus, the anisotropy effect is strong and can plausibly account for the large reflection fraction. 

Furthermore, the corona for Part I has a lower optical depth (though the error bars marginally overlap), with $\tau_{\rm T}= 0.25^{+0.30}_{-0.10}$, than $\tau_{\rm T}= 0.68^{+0.29}_{-0.18}$ for Part II. The scattering in Part I can be predominantly single, in which case the relativistic kinematics forces it to be largely towards the disk. In contrast, there will be more multiple scatterings in Part II, which isotropise the emitted flux. Ideally, all these effects could be incorporated into {\tt comppsc} for the slab geometry, where we would switch between down- and up-emission modes. A development of such an extension of {\tt comppsc} is, however, beyond the scope of this work, where we still use its currently available version. Therefore, while our results demonstrate that the observed strength of the Fe K line {\it can\/} be explained by the two mentioned effects: the curvature of the incident spectrum and the reflection anisotropy, this is not proven owing to our treatment of the latter being only quantitative.

We are aware of the limitations of our treatment of reflection. The {\tt xilconv} model is based on the tables used by {\tt xillver} for the density of $10^{15}$ cm$^{-3}$, while the disk surface density is likely to be much larger. This usually results in a strongly overestimated Fe abundance \citep{Garcia18n}, as in our case. The limited accuracy of {\tt xilconv} was pointed out by \citet{Ding24}. Furthermore, the {\tt xilver} tables have been done for a cold slab heated only by the irradiating photons. In contrast, in the present case, the intrinsic dissipation dominates over heating by irradiation, and the maximum disk blackbody temperature is high, $k T_{\rm max}\approx 1$ keV (which we checked replacing {\tt slimbh} by {\tt diskbb} \citep{Mitsuda84}). However, a more accurate convolution reflection model is not currently available. It would likely have yielded different values for the fitted ionization parameter, the Fe abundance, and the spin. However, it appears that our main result, that the observed Fe K line fluxes can be explained by the irradiation by the spectrum from Compton scattering of disk blackbody photons, is independent of those details, and it would stand. 

The secondary goal is to determine the spin of \source. In our approach, we followed the method of \citet{Parker16}, in which the continuum and reflection methods are combined, and the distance and BH mass are fitted. However, there is a major limitation to this method when using the {\tt kerrbb} model. That model employs the thin-disk model of \citet{NT73} regardless of the Eddington ratio. Therefore, it gives identical spectra for different combinations of $M$, $D$, and the disk mass accretion rate $\dot M_{\rm disk}$ as long as they follow $M\propto D \propto \dot M_{\rm disk}^{1/2}$. However, $L_{\rm disk}/L_{\rm E}\propto D^2/M$ changes. The fit results then differ depending on the chosen initial values.  This is the reason we did not provide these values when reporting the {\tt kerrbb} results\footnote{Therefore, the values of $M$, $D$, and $\dot M_{\rm disk}$ found for GX 339--4 by \citet{Parker16}, who used {\tt kerrbb} are subject to the above scaling, which was pointed out by \citet{Parker19}. The same scaling applies to the {\tt kerrbb} fits in \citet{Zdziarski25a}, where, however, the parameters fitted to other disk models were found preferable. }. Still, the spin and inclination can be obtained with this method. 

On the other hand, {\tt slimbh} spectra depend on the Eddington ratio, based on the slim disk model. Furthermore, they include the radiative transfer calculations of \citet{Davis05, Davis06} as implemented in the {\tt bhspec} model\footnote{\url{https://heasarc.gsfc.nasa.gov/docs/software/xspec/models/bhspec.html}}. Those explicitly depend on both the mass and the Eddington ratio. This breaks the degeneracy present in {\tt kerrbb} and allows us to estimate the mass and distance.

Our results confirm the model dependence of the fitted spins of BH XRBs, as shown using different methods (see \citealt{Zdziarski25a, Zdziarski26}). Here, we obtained $a_*= -0.40^{+0.15}_{-0.23}$ using {\tt kerrbb} and reflection and $a_*= 0.86^{+0.03}_{-0.03}$ using {\tt slimbh} and reflection. At the same time, both results differ from the three previous measurements, yielding $a_*>0.94$ using the reflection method alone. Retrograde accretion in a BH XRB is possible but rare. Outside dense clusters, it requires a BH spin reversal after its formation. \citet{Zdziarski25a} reviewed four published claims of a negative spin in an XRB, and found none of them to be fully convincing.

While the results with {\tt slimbh} appear to provide the currently most accurate estimates, we are aware of their uncertainties. An independent determination of the mass and the distance of \source is highly desirable for testing our model. However, it is tough given the very high extinction toward the source (Section \ref{intro}). 

Another caveat regarding the spin value is the possible presence of a warm and optically thick corona. Such coronae appear to be common in active galactic nuclei; see, e.g., \citet{Petrucci20} and \citet{Ballantyne24}. As we found in Section \ref{joint}, the best fit including this model yields $a_*=0^{+0.08}$, which is similar to the results for a few other BH XRBs (see Section \ref{joint}). Physically, the reduction in the fitted spin is due to the lower maximum temperature of the underlying disk (see figure 6b in \citealt{Zdziarski24b}), with the remaining hard emission accounted for by the warm corona. The temperature reduction leads to an increase in the model value of the disk's inner radius and, consequently, a decrease in the fitted spin. While in \source there is no statistical preference for this model, its presence cannot be ruled out either.

Our obtained spin value is relatively high, with the fitted $a_*\approx 0.8$--0.9. This is much more than most of the spins measured via gravitational waves emitted during mergers of two BHs in binaries, which are $a_*\approx 0.1\pm 0.1$ on average \citep{LVK25b}. Also, it is more than the typical natal BH spin of $a_*\lesssim 0.1$ \citep{Fuller_Ma19, Belczynski20}. An increase of the spin from $a_*=0.1$ to $a_*=0.85$ requires past accretion of about 35\% of the current BH mass, i.e., $\gtrsim 3.5\msun$ (using the formulae of \citealt{Bardeen70}). This can be achieved only in binary evolution models in which donor initial masses in BH LMXBs are significantly higher than those presently observed \citep{Podsiadlowski03, Fragos15}. 

Another caveat for our results is that the standard disk model predicts that the disk is viscously and thermally unstable when dominated by radiation pressure \citep{Lightman74, Shakura76}. This is contrary to observations of the X-ray emission in the soft states of BH XRBs showing their disks to be stable \citep{GD04}, which is also the case for \source (see Table \ref{rms}). Accretion disks can be stabilized by large-scale magnetic fields, e.g., \citet{Begelman07}, \citet{Sadowski16b}, \citet{Mishra20}. The structure of such disks can differ substantially from that of standard disks, as noted by, e.g., \citet{Lancova19}. In particular, the ISCO radius may not form a boundary for the flow, e.g., \citet{Rule25}. Still, the constancy of the inner radius in the soft state of BH XRBs, observed in many sources \citep{GD04}, supports an important role for the ISCO radius. 

While the disk is stable, the corona shows strong variability, see Table \ref{rms}. This behavior of `stable disk/unstable corona' is typical for the soft state of BH XRBs, as first found by \citet{CGR01}, and later in several sources, e.g., Swift 1727.8--1613 \citep{Chand26}. It indicates strongly variable energy deposition in the corona, possibly related to magnetic buoyancy or reconnection. In our models, we have considered the average spectra of both the corona emission and the disk reflection. Since they are expected to be strongly correlated, this approach should not introduce a bias.

We have obtained a relatively large Eddington ratio, $L_{\rm disk}/L_{\rm E}\approx 0.25$ (see Table \ref{fits}), which shows the need to use a slim-disk model. The observed bolometric fluxes are 4.1 and $4.5\times 10^{-8}$ erg cm$^{-2}$ s$^{-1}$, respectively. However, the source is strongly absorbed, and the unabsorbed bolometric flux is about four times larger, $F_{\rm bol}=1.7$ and $1.8\times 10^{-7}$ erg cm$^{-2}$ s$^{-1}$, respectively. The flux in the disk alone (before scattering) is $\approx\! 1.4\times 10^{-7}$ erg cm$^{-2}$ s$^{-1}$. At the best-fit of the distance and the inclination, the total unabsorbed fluxes correspond to the luminosity of $\approx\! 3.5\times 10^{38}$ erg s$^{-1}$, assuming the cosine dependence of the flux ($L=2\pi D^2 F_{\rm bol}/\cos i$). 

\section{Conclusions}
\label{conclusions}

We have modeled the strong and broad Fe K line in \source by reflection of the irradiation of the accretion disk by the spectrum from Comptonization of the disk blackbody by coronal relativistic electrons. We have taken into account the detailed shape of this spectrum, including the crucial effect of the spectral curvature, which reflects the shape of the disk blackbody (see Figure \ref{spectra}). Then, that flux is twice higher than that corresponding to a fit with a power-law irradiation and reflection, e.g., using the reflection models {\tt relxill} or {\tt reflionx}. We have employed convolution models for both Comptonization and reflection, and we emphasize the necessity of using such models in the soft state of BH XRBs.

The distribution of the coronal electrons is hybrid, with a Maxwellian followed by a power-law tail. In Part I of the observation, a significant part of the coronal electrons was relativistic, which then leads to an enhancement of the flux directed back to the disk with respect to that directed outward \citep{Ghisellini91}. These two effects can plausibly explain the presence of the strong Fe K line, though the enhancement of the downward flux was estimated only qualitatively.

We used two relativistic disk models to measure the BH spin via the continuum-fitting method coupled to the reflection method. Using the standard continuum model {\tt kerrbb}, we have obtained a negative spin, $a_*=-0.40^{+0.15}_{-0.23}$. Then, we used {\tt slimbh}, which accounts for both a finite disk scale height at Eddington ratios near unity and the vertical radiative transfer of disk emission. That model gave us $a_*= 0.86^{+0.03}_{-0.03}$, which we consider more likely (though still bearing a significant systematic error) than the former. However, this underscores the strong model dependence inherent in fitting the spins of BH XRBs, and we consider that value tentative. With the {\tt slimbh} model, we estimated the BH mass as $10.7_{-0.9}^{+1.5}\msun$, the distance as $5.1^{+1.1}_{-0.4}$ kpc, and the disk inclination of $36^{+2}_{-2}\degr$. The source Eddington ratio is $\sim$0.25. 

Our preferred spin value may result from accretion of a relatively large mass, $\sim\! 3\msun$, onto the BH during stellar evolution, which could be explained by the model with initial mass masses of donors in LMXBs significantly higher than those observed \citep{Podsiadlowski03}. 

\section*{Acknowledgements}
We thank Jorge Casares for a discussion on the donor of \source, and the two referees for valuable comments. We acknowledge support from the Polish National Science Center grants 2019/35/B/ST9/03944 and 2023/48/Q/ST9/00138. MS acknowledges support from the National Science Center grant 2023/50/A/ST9/00527. BY is supported by the Natural Science Foundation of China (NSFC) grants 12322307, 12361131579, and 12273026, and by the Xiaomi Foundation/Xiaomi Young Talents Program. MP acknowledges support from the JSPS Postdoctoral Fellowship for Research in Japan, grant number P24712, as well as the JSPS Grants-in-Aid for Scientific Research-KAKENHI, grant number J24KF0244.

\bibliographystyle{aasjournal}
\bibliography{../allbib} 

\end{document}